# A Comparative Analysis of E-Scooter and E-Bike Usage Patterns: Findings from the City of Austin, TX

**Mohammed Hamad Almannaa**
Assistant professor, Civil Engineering Department
King Saud University, Riyadh, Saudi Arabia
malmannaa@ksu.edu.sa

**Huthaifa I. Ashqar (Corresponding Author)**
Booz Allen Hamilton
Washington, D.C., 20003 USA
hiashqar@vt.edu

**Mohammed Elhenawy**
Research fellow at Centre for Accident Research and Road Safety
Queensland University of Technology, Australia
mohammed.elhenawy@qut.edu.au

**Mahmoud Masoud**
Research associate at the School of Mathematical Sciences
Queensland University of Technology, Australia
mahmoud.masoud@qut.edu.au

**Andry Rakotonirainy**
Director of the Centre for Accident Research and Road Safety
Queensland University of Technology, Australia
r.andry@qut.edu.au

**Hesham Rakha, PhD, P.Eng**

ORCiD: https://orcid.org/0000-0002-5845-2929

Samuel Reynolds Pritchard Professor of Engineering and Director of the Center for Sustainable

Mobility

Virginia Tech Transportation Institute, Blacksburg, Virginia 24060

Email: hrakha@vt.edu

**ABSTRACT**

E-scooter-sharing and e-bike-sharing systems are accommodating and easing the increased traffic in dense cities and are expanding considerably. However, these new micro-mobility transportation modes raise numerous operational and safety concerns. This study analyzes e-scooter and dockless e-bike sharing system user behavior. We investigate how average trip speed change depending on the day of the week and the time of the day. We used a dataset from the city of Austin, TX from December 2018 to May 2019. Our results generally show that the trip average speed for e-bikes ranges between 3.01 and 3.44 m/s, which is higher than that for e-scooters (2.19 to 2.78 m/s). Results also show a similar usage pattern for the average speed of e-bikes and e-scooters throughout the days of the week and a different usage pattern for the average speed of e-bikes and e-scooters over the hours of the day. We found that users tend to ride e-bikes and e-scooters with a slower average speed for recreational purposes compared to when they are ridden for commuting purposes. This study is a building block in this field, which serves as a first of its kind, and sheds the light of significant new understanding of this emerging class of shared-road users.

**INTRODUCTION**

Due to the large increase in vehicles on the road over the years, cities face challenges in providing high-quality transportation services. Traffic jams are a clear sign that cities are overwhelmed, and that current transportations networks and systems cannot accommodate the current demand without a change in policy, infrastructure, transportation modes, and commuters' choice of transportation mode. Previous research have shown that drivers are more likely to use their personal cars or even the ridesharing services very often for short-distance trips, compared to long ones [1]. For example, Uber's ride lengths (and private car trips) in the US are skewed towards the low end, which means that Uber rides are more likely to be short (< 5 mi). Consequently, micro-transport modes such as e-scooter-sharing systems will replace about 50 percent of Uber trips.

As an effort to support micro-transportation modes such as bicycling and offering alternative transportation modes, Bike-sharing dock-based systems (BSSs) were implemented in more than 50 countries [2]. Through distributing bicycles from stations distributed across a service area, BSSs seek to inspire people to travel through bike. People can borrow and return a bicycle from any station around their destination. Bicycles are considered an affordable, easy-to-use, and, healthy transportation mode, and BSSs show significant transportation, environment, and health benefits.

In transportation, BSSs replace privately-owned car trips with bicycling, thereby mitigating traffic jams in the city. A survey conducted by McNeil at al. found that 80% or more of BSS users said they use BSSs for shopping/errands, social/recreational, trips to and from public transit, and commute trips [3], confirming that BSSs are becoming a reliable and convenient transportation mode for both recreational and non-recreational trips. In the environmental and health fields, the reduction in privately-owned car trips means less carbon energy consumption and carbon emissions. Qiu and He found that using BSSs in Beijing could save workers 8 minutes per day and that this saving could result in reducing fuel consumption by 225.05 thousand tons [4]. This would contribute in increasing the GDP of Beijing by Ren Min Bi (RMB) 1.2 billion (RMB is the official currency of china) and reducing the health costs by RMB 2420.57 million yuan.

As the use of dock-based BSSs has grown, imbalance has become an issue and an obstacle for further growth. Imbalance occurs when bikers cannot drop off or pick-up a bike because the bike station is either full or empty [5-9]. To overcome the limitation of the dock-based BSS, a new generation of BSSs was introduced in China in 2015—the dock-less (or station-free) BSS takes an approach in which the BSS does not have stations. Rather, bikes (or e-bikes) are distributed along city sidewalks. People can rent a bike from anywhere and leave it within a defined zone instead of certain location ("stations").

In 2012, a modern mode of transportation (micromobility); the e-scooter network debuted along with (and sometimes competitively) the dockless bike-sharing system and thrived in 2017. This system works similarly as the bike-sharing system, yet with e-scooter instead of a bike. With more than two million users, the e-scooter sharing system has entered more than 80 city schemes since 2017 [10]. In 2018, the e-scooter system has exceeded both dock-less and dock-based bike sharing system as shown in FIGURE 1. This increased usage of e-scooter system compared to other public transportation systems has gained policymakers and investors' attention to investigate so that they better ensure safety, management and operation.

Limited regulation and lack of awareness of the risks involved in using these micro-mobility transportation modes raises numerous safety concerns. For example, a U.S. national database showed that facial and head injuries from micro-mobility devices have tripled over the last 10 years [11]. This study found that the number of incidents climbed from 2,325 in 2008 to 6,957 in 2018, 66% of those treated were not wearing helmets. Another study in Los Angeles and Santa Monica found that e-scooters have been associated with 249 emergency room visits between September 1, 2017 and the end of August 2018 [12]. In this paper, we investigate the users' behavior of e-scooter and dock-less e-bike sharing system. We investigate how average trip speed change depending on the day of the week and the time of the day using publicly available dataset collected in the city of Austin in Texas State, U.S. for the years of 2018-2019.

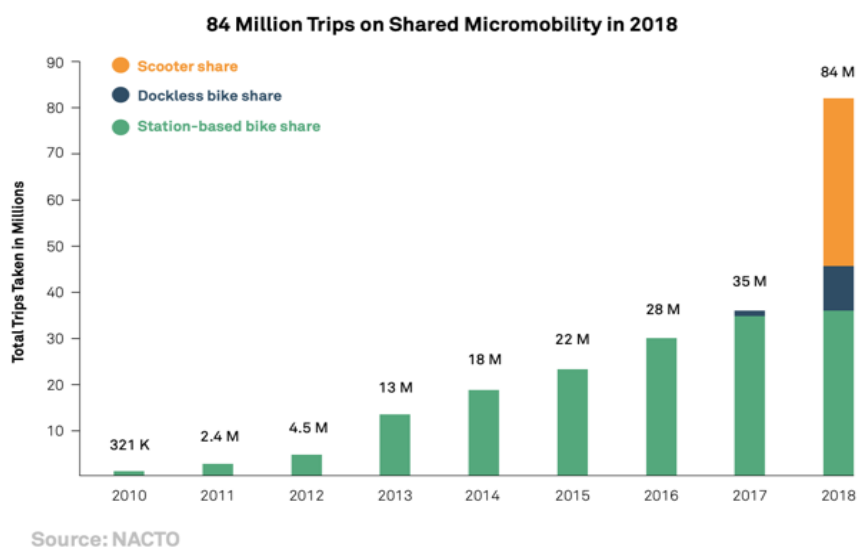

**FIGURE 1  Trips taken by station-based, scooter-based, and dockless systems.**

## RELATED WORK

Previous research has focused on investigating the demographic and behavior of BSS users using either surveys or analyzing real data [13, 14]. Buck et al., for example, found that males use BSSs 30% more than females, yet their survey showed the distribution of males-to-females to be equal (48–52%) [15]. In the UK, a survey was conducted to investigate BSSs' users. Results showed the same conclusion; that men use BSSs more than women but by 17%. Other research efforts were conducted to study the differences between long-term and short-term users of BSSs. For example, in Washington, D.C., a survey showed that 43% of long-term users were work commuters while short-term users tended to have fewer work trips [15]. Also, in London, UK, a survey revealed the same conclusion; that long-term or annual members seems to be work commuters.

Other researchers have looked at the purpose of the trips for long-term BSSs users [16]. Raviv and Kolka found that 52% of BSS long-term users used BSSs for work in the last trip [17]. Similar, another survey conducted in Brisbane, Australia showed the same results that the main purpose of the last trip was leisure for short-term members but work-related for long-term users [18]. However, a study in 2013 shows that short- versus long-term use depended on many factors, such as gender, age, residential location, ethnicity [18], and weather [19, 20].

Buck et al. studied the demographic of BSS users and traditional cyclists in Washington, D.C. [15]. The authors found the user demographic to be mostly white, around 80% for both traditional and BSS cyclists. Compared to the overall Washington D.C. population, the population of black people is slightly higher than white people (by 5%). However, none of the aforementioned research efforts has conducted temporal analysis of BSSs such as the speed of the bikers with respect to a time event.

On the other hand, there has been a number of studies concerning e-scooter-sharing systems [21-24]. Hardt and Bogenberger conducted a real-field experiment a pilot study in the city of Munich, Germany to determine if e-scooter-sharing system could replace vehicles [25]. Their results show that e-scooters could replace most of the vehicle trips. Also, Smith and Schwieterman have tested the possibility that e-scooter sharing system supports the public transportation service in Chicago [23]. They found that e-scooter-sharing system is a strong alternative for private vehicles for short distances, and thus has the potential to increase the number of car-free households in Chicago.

Nocerino et al. analyzed and tested the performance of both electric bicycles and electric scooters for delivering goods in seven European countries. They mainly investigated the possibility of replacing the traditional freight transport service and reducing $CO_2$ emissions and energy savings in urban areas. Degele et al. studied the pattern usage of e-scooter in Germany. They concluded that the use of e-scooter sharing system can be grouped into four segments based on: age, time between rides, distance driven, and revenue per customer. They categorized into four groups: demographic, geographic, behavior, and psychographic. However, none of the above research efforts has considered comparing the behavior of e-scooter to bike sharing systems users.

A few studies have been conducted to compare the pattern usage of e-scooter to dock-based bike sharing systems [26-28]. McKenzie conducted a spatiotemporal analysis of e-scooter-sharing and dock-based bike-sharing usage patterns in Washington, DC. He found there is a remarkable difference between the two systems in terms of the usage. One of which is bikes are used to commute for work while e-scooters are not. However, the comparison here is not fair given the bike-sharing system requires users to leave bikes in certain locations ("stations") unlike e-scooter-sharing systems. Fawcett et al. attempted to analyze the e-bicycle and e-scooter-sharing systems with a focus on safety risks [27]. They suggested tips for operators and investors on how they can manage to overcome these risks. However, they did not study and analyze the users' behavior of both systems with respect to the operational aspect: temporal analysis.

## DATA SET

This study used a dataset that was collected in the city of Austin, Texas and is publicly available by the City of Austin [29]. The dataset contains about 6 million trips taken by users for either e-bikes or e-scooters from December 3, 2018 to May 20, 2019. Each trip is represented in a row and each of which has 18 features: trip ID, device ID, vehicle type (e-scooter or e-bike), trip duration (in seconds), trip distance (in meter), start time of the trip, end time of the trip, month, hour, day of week, year, council district (both start and end), and census tract (both start and end). In our analysis, we included the trips that meets the following criteria: (1) trip distance greater than or equal to .1 miles and less than 500 miles, and (2) trip duration less than 24 hours. TABLE 1 shows the total number of e-scooter and e-bikes that operators in the city.

**TABLE 1 List of operators licensed to serve in Austin.**

| Operator | Scooters | Bikes |
|----------|----------|-------|
| Bird | 4500 | 0 |
| JUMP | 2500 | 2000 |
| Lime | 5000 | 0 |
| Lyft | 2000 | 0 |
| OjO | 100 | 0 |
| Skip | 500 | 0 |
| Spin | 500 | 0 |
| VeoRide | 300 | 50 |

**METHODOLOGY**

In [30], we proposed a multi-objective clustering algorithm with a goal of maximizing the purity and similarity in each cluster simultaneously as shown in FIGURE 2. This algorithm was developed based on the CA (College admission) algorithm [31] where it considers the true labeling of the dataset (i.e., weekday, daytime) and simulate the clustering problem as a game. Two disjointed player sets in this game enter the game to define a cohesive match. The set for the first player is the centroids (clusters), and the set for the second player is the data examples (data points). That centroid controls the data points based on the intra-cluster distances in its preferences list. Additionally, that data point ranks the centroids in their preferences list based on their concentration (i.e. high purity). A data point gives centroids a higher preference when most of its members have the same mark as their own. In summary, the proposed algorithm aims to match between clusters that want to minimize intra-cluster distances and maximize purity until it converges. More information can be found in [30].

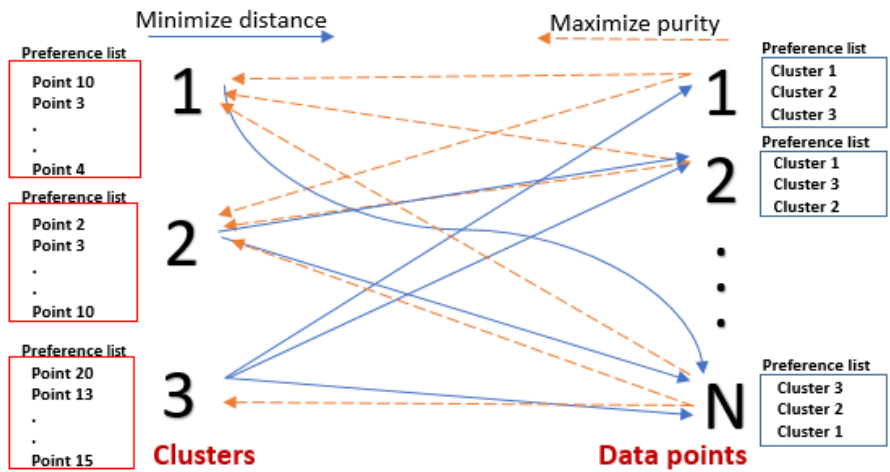

**FIGURE 2 CA based clustering.**

In this research effort, we used this algorithm to understand how average trip speed of micro-mobility transportation modes change depending on the day of the week and time of the day. The consensus clustering (CC) technique was used to find the optimal number of clusters [32], which is a methodology based on resampling that estimates the number of clusters in a dataset by capturing the consensus among several clustering runs. CC attempts to produce data partitions that are more robust than the ones we may expect to obtain by application of a single clustering algorithm to the observed data.

**RESULTS**

Comparing the usage pattern of e-scooters and dockless e-bikes is important to operators and investors so they can ensure their safety and identify the role that these play in a multi-modal transport system. In this study, we analyze an evolving dataset of e-scooter and dockless e-bike usage patterns in Austin, TX from December 3, 2018 to May 20, 2019. For each of the day-of-week and time-of-day, we (1) started the analysis by determining the number of clusters to be used in the dataset using the CC technique; (2) applied the CA-based clustering algorithm to cluster the usage pattern of e-scooters and dockless e-bikes based on the outcome of the first step; and (3) tested the null hypothesis that the usage pattern of the dataset for each case are samples from continuous distributions with equal medians, against the alternative hypothesis that they are not using the Wilcoxon rank-sum test.

**Day-of-week**

Based on the abovementioned steps, FIGURE 3 shows the results of the CC technique. It shows the consensus index value for different clusters for the dockless e-bikes and e-scooters. The flattest curve corresponds to the most stable model order, which is $k = 2$ for e-bikes and e-scooters.

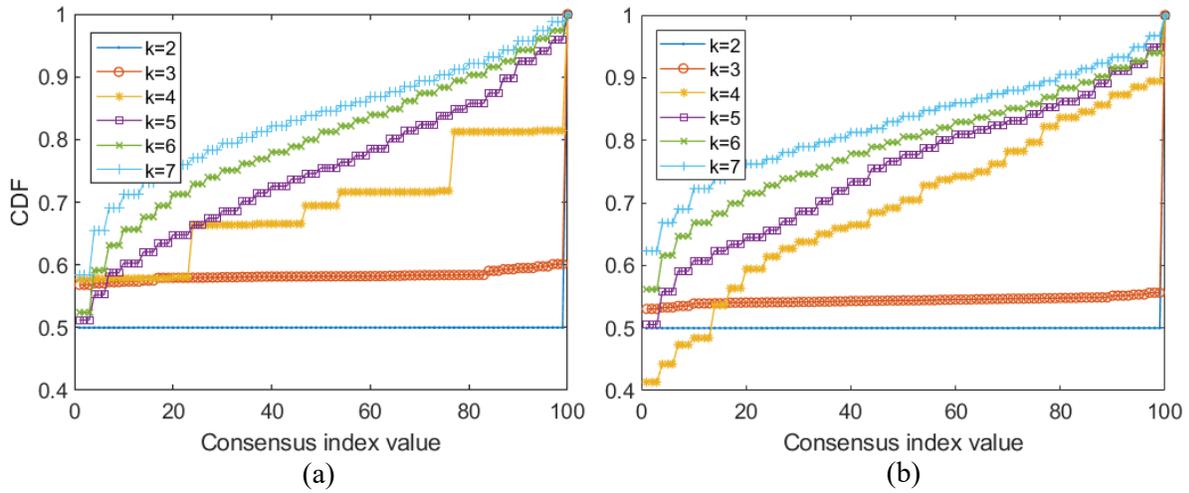

**FIGURE 3** Consensus index value for different clusters for (a) e-bikes, and (b) e-scooters.

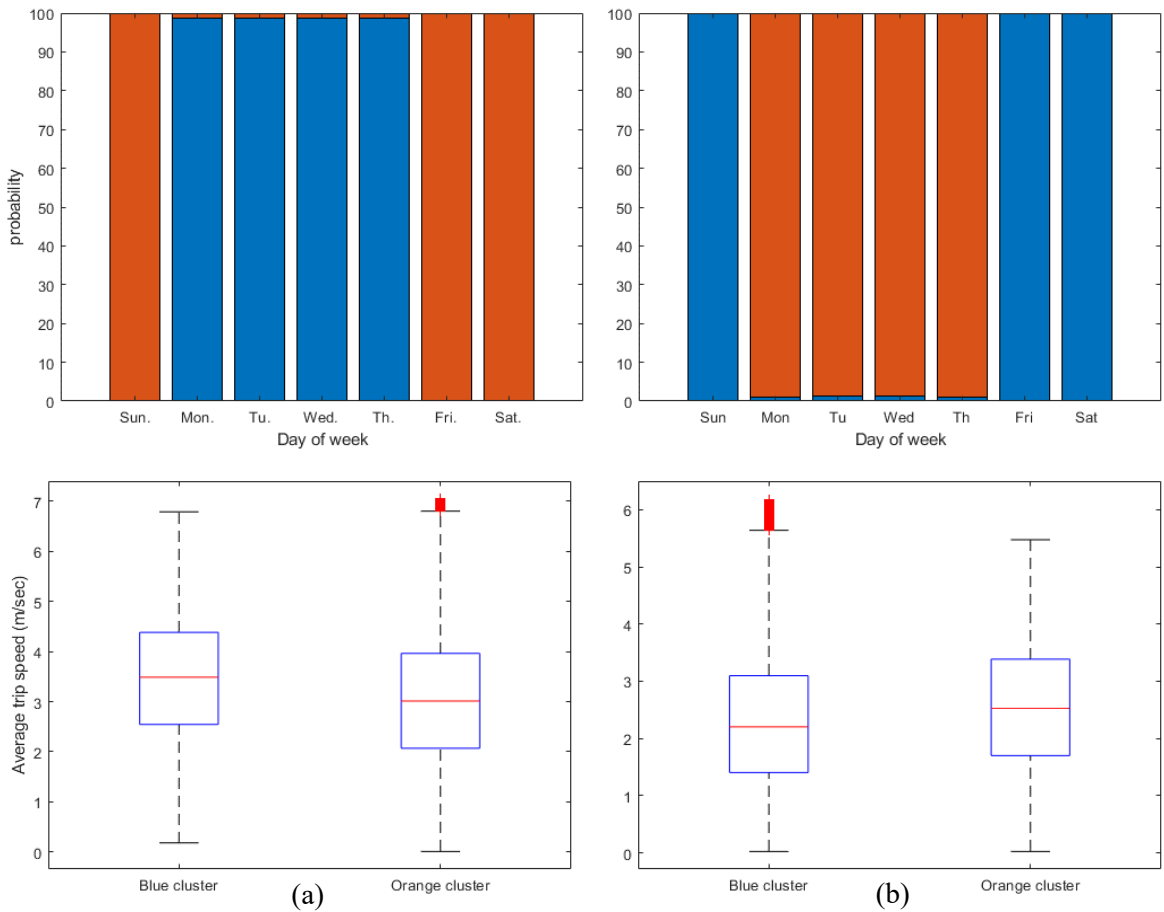

**FIGURE 4 Clusters of day-of-week for (a) e-bikes, and (b) e-scooters.**

**TABLE 2 Mean and standard deviation of each clusters for e-bikes and e-scooters.**

| Cluster Color | Mean ($m/s$) | Std ($m/s$) |
|---|---|---|
| **E-Bikes** | | |
| Orange | 3.01 | 1.38 |
| Blue | 3.44 | 1.31 |
| **E-Scooters** | | |
| Orange | 2.55 | 1.19 |
| Blue | 2.32 | 1.25 |

**Time-of-day**

Similar to the week-of-day analysis, results of the CC technique are shown in FIGURE 5. It shows the consensus index value for different clusters for the dockless e-bikes and e-scooters for the time-of-day data. The flattest curve corresponds to the most stable model order, which is $k = 2$ for e-bikes and e-scooters.

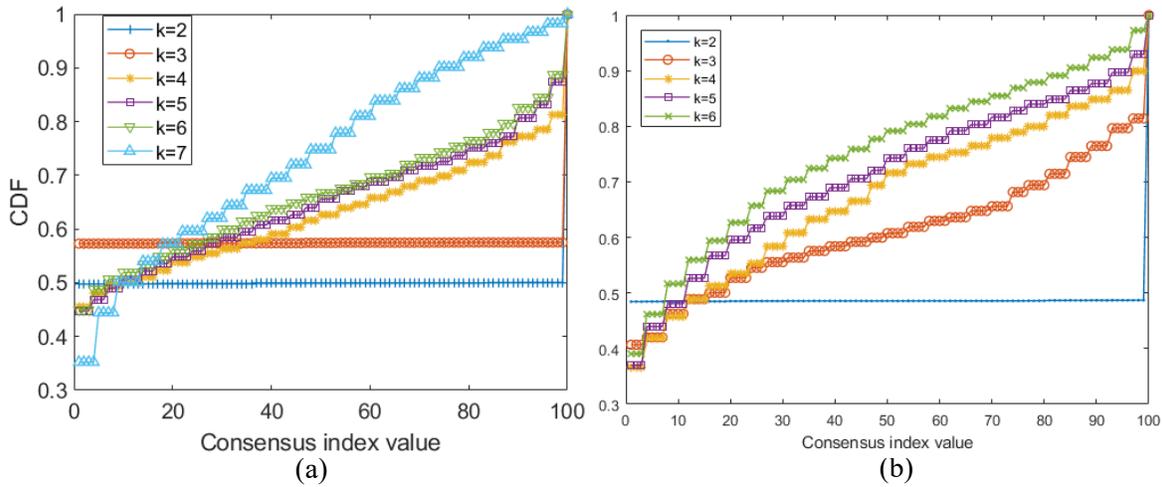

**FIGURE 5 Consensus index value for different clusters for (a) e-bikes, and (b) e-scooters.**

FIGURE 6 shows the results of clustering the time-of-day for dockless e-bikes and e-scooters based on the average trip speed. A two-sided Wilcoxon rank-sum test was then used to test the null hypothesis that data in the two clusters are samples from continuous distributions with equal medians, against the alternative that they are not. We were able to reject the null hypothesis in the case of e-bikes and e-scooters with a p-value of *zero* for both. Results present a different usage pattern in the distribution of the average speed of e-bikes and e-scooters over the hours of the day. For e-bikes, an average speed cluster of $3.32\ m/s$ starts from 12AM midnight and lasts until 11AM, in which another cluster of lower average speed, $3.09\ m/s$, runs the rest of the day except from 6-7PM. For e-scooters, an average speed cluster of $2.78\ m/s$ starts from 3AM and lasts until 12PM and the lower average speed cluster, $2.19\ m/s$, runs the rest of the day. The mean and standard deviation of each cluster for e-bikes and e-scooters is shown in TABLE 3.

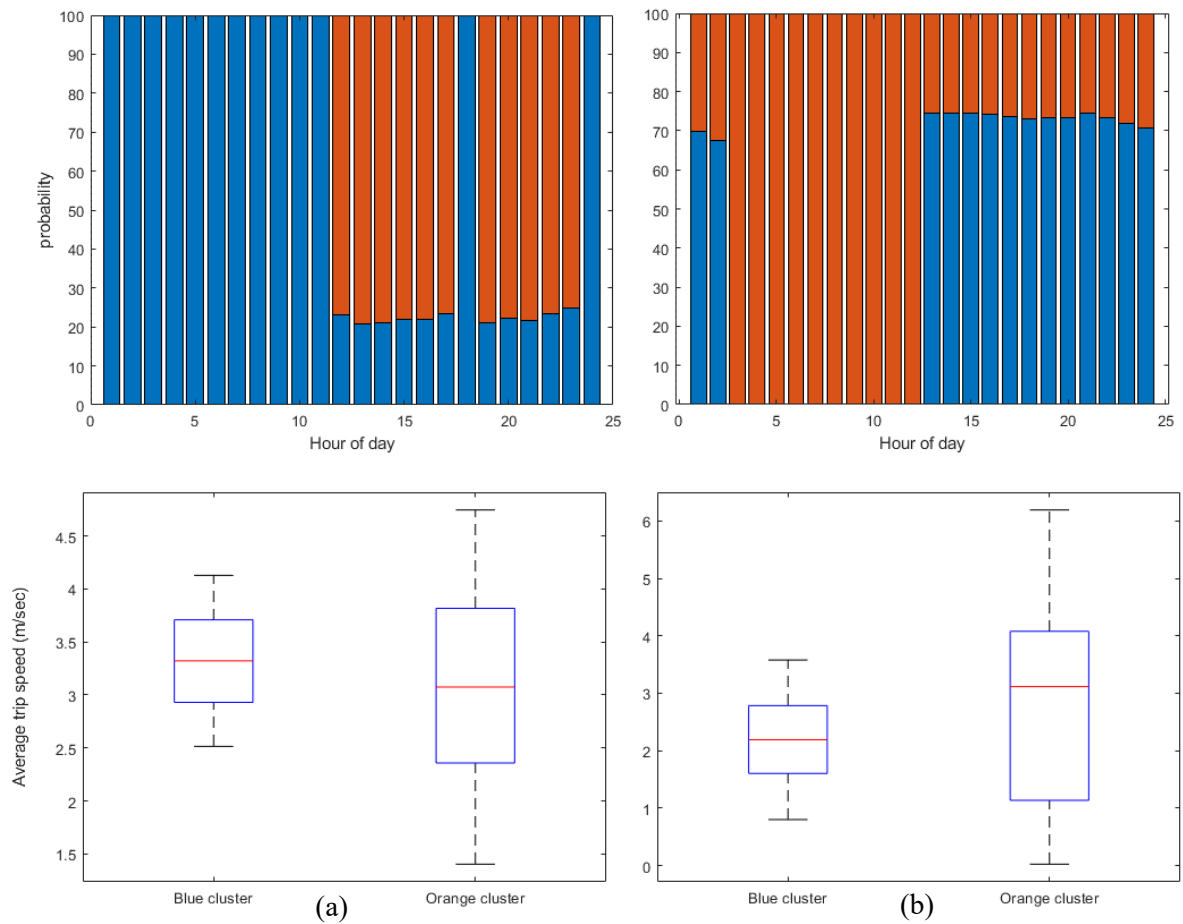

**FIGURE 6 Clusters of time-of-day for (a) e-bikes, and (b) e-scooters.**

**TABLE 3 Mean and standard deviation of each clusters for e-bikes and e-scooters.**

| Cluster Color | Mean ($m/s$) | Std ($m/s$) |
|---|---|---|
| **E-Bikes** | | |
| Orange | 3.09 | 0.89 |
| Blue | 3.32 | 0.46 |
| **E-Scooters** | | |
| Orange | 2.78 | 1.64 |
| Blue | 2.19 | 0.73 |

## DISCUSSION AND CONCLUSIONS

Limited regulation and lack of awareness of the risks involved in using these micro-mobility transportation modes raises numerous safety concerns. The purpose of this study was to investigate the differences of users' usage patterns of e-bikes and e-scooters in sharing systems in the United States, where e-scooters and e-bikes are a relatively emerging transportation mode in many cities and the e-scooters and e-bikes share in the market has changed since few studies in the literature.

Our results show that the trip average speed for e-bikes ($3.01 - 3.44\ m/s$) is higher than the trip average speed for e-scooters ($2.19 - 2.78\ m/s$), in general. Results also show a similar usage pattern of the average speed of e-bikes and e-scooters throughout the days of the week, in which one cluster in the weekdays and the other one in the weekends. However, a different usage pattern was found in clustering the average speed of e-bikes and e-scooters over the hours of the day. On one hand, one e-bikes cluster starts from 12AM midnight and lasts until 11AM and the other cluster runs the rest of the day except from 6-7PM. On the other hand, for e-scooters, a cluster starts from 3AM and lasts until 12PM and the other cluster runs the rest of the day. Furthermore, users tend to ride e-bikes and e-scooters with a slower average speed for recreational purposes (i.e. through weekends and non-working hours), compared to when they ride them for commuting and transporting purposes (i.e. through weekdays and working hours).

There are some ideas that could serve as future steps to this study. First, this study could be implemented to data from different geographical locations in the U.S., as this was the only data available of this kind to the best of our knowledge. A different dataset from different locations in U.S. could further leverage this study to increase its veracity. We presume that, as e-bikes and e-scooters penetrate deeper into different locations and if dataset that contains trips for the different seasons and from different locations will be investigated; the usage patterns might shift. Still, the results of this study are primarily contains significant finding in this field to build on, the first of its kind, and sheds the light of significant new understanding of this emerging class of shared-road users that policy-makers, operators, and researchers would find intriguing to understand for safety, transportation, health, and recreation purposes.

## DATA AVAILABILITY

The dataset used to support the findings of this study is publicly available through https://austintexas.gov/micromobility.